\documentclass[journal]{IEEEtran}

\usepackage{cite}

\ifCLASSINFOpdf
  \usepackage[pdftex]{graphicx}
  \graphicspath{{./figs/}}
  \DeclareGraphicsExtensions{.pdf,.jpeg,.jpg,.png}
\else
\fi

\usepackage{subcaption}

\usepackage{array}
\usepackage{url}

\begin{document}
\title{Potential Features of ICU Admission in X-ray Images of COVID-19 Patients}
%
%
%

\author{Douglas~P.~S.~Gomes,
        Anwaar~Ulhaq,
        Manoranjan~Paul,
        Michael~J.~Horry,
        Subrata~Chakraborty,
        Manash~Saha,
        Tanmoy~Debnath,
        D.M.~Motiur~Rahaman
\thanks{D.P.S. Gomes, Anwaar Ulhaq, Manoranjan Paul, and Tanmoy  Debnath were with the Machine Vision and Digital Health (MAVIDH) Research group, School of Computing and Mathematics, Charles Sturt University, NSW, Australia. E-mails: douglas.uf@gmail.com, aulhaq@csu.edu.au, mpaul@csu.edu.au, tdebnath@csu.edu.au.}
\thanks{Michael J. Horry and Subrata Chakraborty were with the Centre for Advanced Modelling and Geospatial Information Systems (CAMGIS), School of Information, Systems, and Modeling, Faculty of Engineering and IT, University of Technology Sydney, NSW, Australia. E-mails: michael.j.horry@student.uts.edu.au, subrata.chakraborty@uts.edu.au}
\thanks{Manash Saha was with the Wollongong Hospital, NSW, Australia. Email: manash.saha@health.nsw.gov.au}
\thanks{DM Motiur Rahaman was with the National Wine and Grape Industry Centre, Charles Sturt University, Australia. Email: drahaman@csu.edu.au.}
\thanks{Michael J. Horry was also with the IBM Australia Limited, Sydney, NSW, Australia.}
}


\maketitle

\begin{abstract}
X-ray images may present non-trivial features with predictive information of patients that develop severe symptoms of COVID-19. 
If true, this hypothesis may have practical value in allocating resources to particular patients while using a relatively inexpensive imaging technique. 
The difficulty of testing such a hypothesis comes from the need for large sets of labelled data, which need to be well-annotated and should contemplate the post-imaging severity outcome. 
This paper presents an original methodology for extracting semantic features that correlate to severity from a data set with patient ICU admission labels through interpretable models.
The methodology employs a neural network trained to recognise lung pathologies to extract the semantic features, which are then analysed with low-complexity models to limit overfitting while increasing interpretability. 
This analysis points out that only a few features explain most of the variance between patients that developed severe symptoms.
When applied to an unrelated larger data set with pathology-related clinical notes, the method has shown to be capable of selecting images for the learned features, which could translate some information about their common locations in the lung.
Besides attesting separability on patients that eventually develop severe symptoms, the proposed methods represent a statistical approach highlighting the importance of features related to ICU admission that may have been only qualitatively reported.
While handling limited data sets, notable methodological aspects are adopted, such as presenting a state-of-the-art lung segmentation network and the use of low-complexity models to avoid overfitting.
The code for methodology and experiments is also avaliable \footnote{\url{https://github.com/dougpsg/covid_mavidh_icufeatures_scoring}}. 

\end{abstract}


%
\IEEEpeerreviewmaketitle

\section{Introduction}
\IEEEPARstart{C}{OVID-19} remains a serious worldwide health pandemic with infections numbering 21,294,845 as of 16 August 2020 with 761,779 deaths \cite{world2020coronavirus}. One emerging characteristic of COVID-19 disease is the wide range of symptoms experienced by infected persons ranging from entirely asymptomatic through admission to the general ward for a range of symptoms including fever, cough, fatigue, headache and diarrhoea \cite{huang2020clinical} to severe pneumonia requiring admission to ICU with mechanical ventilation \cite{guan2020clinical}. Reported case-fatality rates vary from 1\% to greater than 7\%, usually due to respiratory failure \cite{vincent2020understanding}.  

Whilst care in modern ICUs will result in death rates towards the lower end of the range, life-sustaining therapies will, in practice, be limited by lack of personnel, infrastructure and materials and equipment. This limitation of resources leads clinicians to make prognostic decisions based on criterion such as old age, fragility and comorbidity that can lead to the death of patients with poor prognosis in favour of patients with better prognosis \cite{vincent2020understanding}. 

The most widely used testing methodology for COVID-19 is the real-time reverse-transcriptase-polymerase-chain-reaction (RT-PCR) test \cite{huang2020clinical,tang2020laboratory}. This test provides a binary indication of whether a person is infected with COVID-19 or not. Similarly, serology point of care tests that detect COVID-19 antibodies provide a retrospective measure of infection \cite{tang2020laboratory} but do not provide information relating to the severity or progression of the disease and are not useful in helping clinicians to devise an individual plan of treatment for a patient. 



Chest medical imaging has proven to be useful in managing more serious COVID-19 infections since respiratory dysfunction iso ne of the primary sources of COVID-19 morbidity and mortality. 
Several researchers have shown Deep Learning to be useful in classifying COVID-19 cases from medical images including CXR \cite{civit2020deep,oh2020deep,ucar2020covidiagnosis,yoo2020deep}, CT \cite{li2020community,ni2020deep,zhang2020clinically} with some research also achieving promising results with the Ultrasound imaging mode \cite{born2020pocovid,roy2020deep,horry2020covid}.
However, using Deep Learning as a diagnostic tool can be problematic as it is hard to assess biases, risk, potential overfitting, and ability to generalize in clinical settings \cite{rubin2020role, wynants2020prediction}.
Its value resides more on the prognosis and treatment side than in actual diagnostic use \cite{cohen2020covidProspective}.
Chest X-rays (CXR) and Computed Tomography (CT) imaging are useful tools in the management of COVID-19 cases since these methods help clinicians to establish a baseline pulmonary status and identify underlying pulmonary conditions that may contribute to the patients’ risk. 
Chest X-rays, in particular, are also non-invasive and can be potentially used as bed-side patient monitoring tool \cite{cohen2020predicting}.
Compared to CT, CXR is less expensive, more available, and require less technical expertise to perform and interpret than Ultrasound \cite{ulhaq2020review}. 
Over the course of the COVID-19, subsequent imaging can be useful for assessing COVID-19 progression and secondary complications \cite{rubin2020role}.

This paper therefore focuses on the potential disease progression and provides a machine learning-based method for assessing features with predictive potential to classify patients who are more likely to develop severe COVID-19 symptoms and be eventually admitted in ICU. 
The method proposed here leverages the existence of a data set with patient outcome labels (only publicly accessible for CXR images) and the fact that such images were taken before the patients were admitted to ICU. 
As such a data set has a small sample size, significant measures were taken to aggressively limit potential overfitting; from image pre-processing techniques and lung segmentation to model and feature selection. 
Measures like lung segmentation and cropping are not only needed given the data set size but also to prevent the method to rely on confounding image features, as already pointed as a known problem in a recent work \cite{degrave2020ai}.
For these tasks, lung segmentation and cropping, a model achieving state-of-the-art results is also proposed and part of the contribution herein.
The validation is not only attested from separability metrics using the selected semantic features in the ICU data set but also by correlating the results to a more extensive external data set.
Despite not having ICU-related labels, such a data set contains pathologies and localizations labels in its metadata that allows for assessing the significance of the selected semantic features. 
The authors present such a machine learning system in the hopes that it will inform practitioners of the weight of particular features in disease progression, thereby helping early intervention to improve patient outcomes and plan better to allocate critical resources.
A second significant benefit of this novel framework is that it can be used as a method to assist the medical community in quickly identifying features in medical images that are associated with severe progression of COVID-19, similar diseases, or other potential future infections. 

\section{Methods}
\begin{figure}[!t]
\centering
\includegraphics[width=\columnwidth]{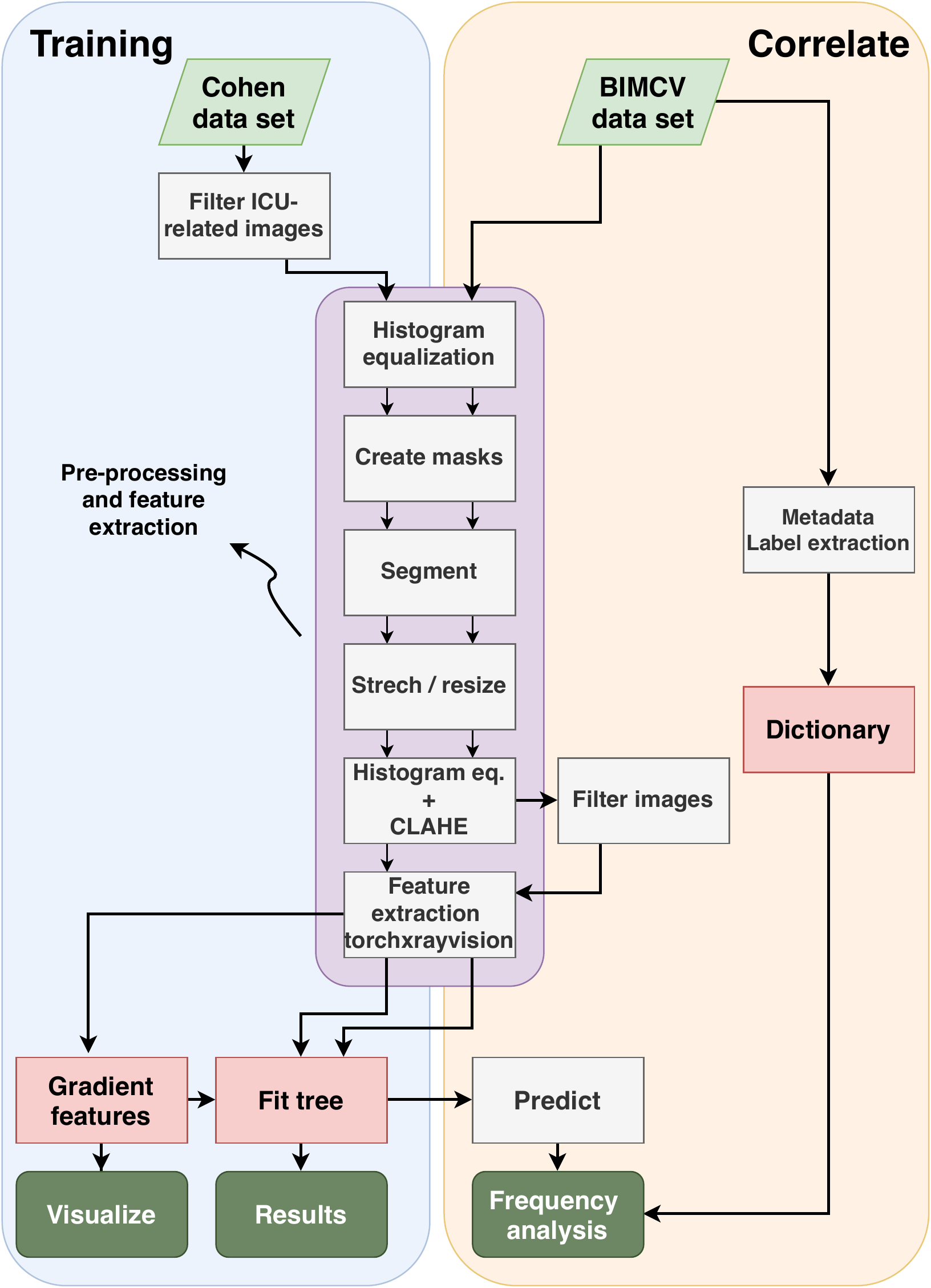}
\caption{Graphical summary of the methodology. ICU-related images are filtered from a limited data set, processed, and features are extracted via a specialized neural network. A few of the extracted features are used to fit a shallow decision tree, which is further applied to an external data set. Results from training and external validation are compared, and visualizations are presented.}
\label{methodology_diagram}
\end{figure}

Given the limited amount of data to learn the severity-correlating features, a focus on limiting potential overfitting was central to most methodological decisions. Instead of disregarding such investigation simply because the data is limited, the idea proposed here is to use low-complexity, interpretable methods to test features for potential predictive value. Therefore, in an attempt to increase the features’ generalization potential, a conservative bias for complexity was adopted when choosing models and hyper-parameters. Some of this concern for overfitting manifested as procedures adopted in the pre-processing of images: histogram equalizations (adaptive and standard) and lung segmentation and cropping. In the learning stage, more importantly, it inspired the choice to use a low complexity (low Vapnik–Chervonenkis dimension), shallow decision tree, as to limit the ability of the method to simply shatter the data set. Such an approach also presents the considerable advantage of resulting in a highly interpretable model, which is much desired in systems designed to support human decision-making. To improve understandability, Fig. \ref{methodology_diagram} is set to depict a high-level graphical summary of the methodology. 

The authors nevertheless understand that, despite the care taken for overfitting and model selection, it is highly desirable to have additional validation methods that can attest for the model's generalization ability.
To this end, this study also contemplates the application of such a model to a larger, different dataset (1000+ images). Although lacking the metadata regarding ICU admission, such X-ray images have clinical annotations that can be tokenized and correlated to the semantic features learned by the low-complexity model. Given the features’ semantic nature, one can also correlate them to other works in the related literature that report on the lung pathologies present in patients that developed severe symptoms.

\subsection{Setting up the ICU data set}
The data set used to learn the features present in patients with higher chances of being admitted to ICU was a subset of the data presented by Cohen et al. \cite{cohen2020covid}. It is one of the most popular data sets on the literature, favourited more than 2000 times on Github. One of its most positive characteristics is its rich metadata containing categories such as sex, age, location, patient condition, and outcome (ICU admission), allowing for the investigation presented here. However, it should be noted that many of such interesting labels are sparsely distributed in the data set and not present for every every image. For the images that had rich descriptors, two prominent labels were of interest when setting up this ICU data set: ‘went-icu’ and ‘in-icu’. An image taken from a patient marked with ‘Y’ on the former label and ‘N’ on the latter is a sample from a patient that eventually developed severe symptoms before they were admitted in ICU. Therefore, one can reasonably form the hypothesis that there might be features in these images that are associated with patients that were eventually admitted in ICU. 
In total, 100 images containing these two labels were used in the analysis, which were further multiplied by a factor of approximately 10 (1040 images) by gentle affine random data augmentations: rotation, piecewise affine transformation, translation, and shear.

\subsection{Pre-processing pipeline}
The selected images are assigned to a class (ICU or not) and fed through a pre-processing pipeline established to normalize and remove potential bias-inducing artifacts. The sequential blocks, illustrated in Fig. \ref{methodology_diagram}, are responsible for histogram equalizations, lung segmentation and lung-area cropping. Histogram equalization is set to impose a normalizing effect on the contrast of images by equalizing the distribution of pixel intensities that might be concentrated in a narrow range. Such a procedure usually improves the visual contrast in the X-ray images, which can come from particular distribution depending on the equipment manufacturer.   

Subsequent to the normalization step, a lung-segmentation model is applied to each X-ray image to improve the signal-to-noise ratio for machine learning classification.
Such a step is potentially the most important regarding the mitigation of potential bias as all areas that are not lung-related are excluded, allowing the model the focus on lung-specific features.
Regarding COVID-19 as a pulmonary disease, the objective region of interest considered when segmenting the lung was the region within the thorax, primarily the lung lobes.
The utility of this approach in medical image classification is supported in past studies \cite{rabinovich2007does}, and recent works have pointed the fact that that lack of such practice can make the model focus on confounding, not disease-related features \cite{degrave2020ai}. 
A comprehensive review of lung area segmentation techniques may be found in \cite{candemir2019review}, which notes that Deep Learning methods provide similar accuracy as in inter-observer performance at the expense of a long training process.  

\begin{figure}[t]
  \centering
  \begin{subfigure}[t]{0.3\columnwidth}
    \includegraphics[width=\textwidth]{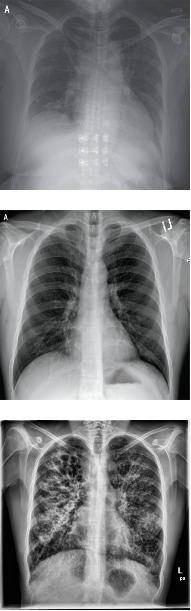}
    \caption{}
  \end{subfigure}             
  \begin{subfigure}[t]{0.3\columnwidth}
    \includegraphics[width=\textwidth]{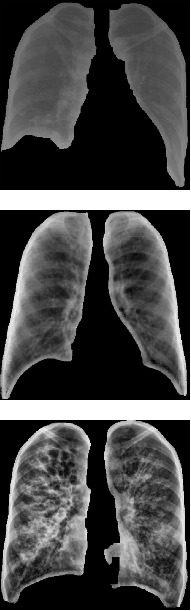}
    \caption{}
  \end{subfigure}
  \begin{subfigure}[t]{0.3\columnwidth}
    \includegraphics[width=\textwidth]{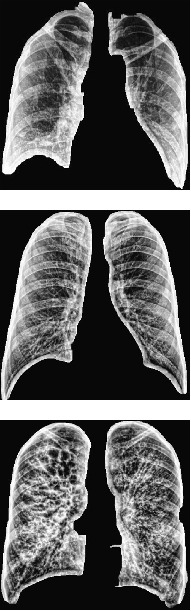}
    \caption{}
  \end{subfigure}
  \caption{Chest X-Ray images in different stages of the pre-processing pipeline. a) Original images with different contrast and histogram. b) Segmented lungs without the normalization techniques. c) Images with lungs segmented and normalization applied.}
  \label{image_proc_pipeline}
\end{figure}

With the assumption that one could improve the U-Net architecture to yield better dice similarity co-efficient results, as well as using training data from two additional data sources, a lung-segmentation model was proposed and trained. The additional images are from Montgomery and Shenzen datasets \cite{jaeger2014two} and resulted in a combination corpus of 1185 CXR image/mask pairs. As typical U-net designs are based on simple convolution stacks similar to the VGG network architecture \cite{simonyan2014very}, a U-Net based on a skip-connection architecture following \cite{Mique_2020,zhang2018road} is also proposed. Such an architecture allows the network to re-use features learned in earlier layers, thereby improving its ability to learn the identity function and increase generalization. As an additional benefit, this architecture also allows the U-Net to train on a smaller number of epochs, thereby reducing the training time.  
Given the state-of-the-art dice similarity coefficient achieved by the lung-segmentation model proposed here, validated in three independent data corpora, it is expected that the model will be useful in unseen images. Artefacts in the generated lung field masks were removed by a combination of morphological closing, contour filling and flood-filling, resulting in a set of automatically segmented lung-field images. As a final step, the images were also automatically cropped to fully contain the segmented lung field, resulting in a uniform image set to the downstream classifier and normalizing for different lung sizes. In examples of the result of applying the pre-processing pipeline, Fig. \ref{image_proc_pipeline} illustrates three images in the input and output stage, with and without normalization. 

\subsection{Feature extraction}

\subsubsection{TorchXRayVision}
The feature extraction model is a DenseNet \cite{huang2017densely} pre-trained on 80,000+ lung X-ray images. The model, part of the TorchXRayVision library, was firstly presented in \cite{cohen2020limits} and also later used in a work proposing a COVID-19 pneumonia severity score \cite{cohen2020predicting}. A particular aspect of this model is that the author trained it to detect specific lung pathologies from large data sets such as the CheXpert (Stanford) \cite{irvin2019chexpert}, ChestX-ray8 (NIH) \cite{wang2017chestx} and MIMIC-CXR (MIT) \cite{johnson2019mimic}. The semantic labelling was performed by adding an 18-node layer at the end of the network and training it to classify different pathology labels through a sigmoid activation layer. The labels that each of the nodes was trained to classify were the ones found in the large data sets used: \emph{Atelectasis, Cardiomegaly, Consolidation, Edema, Effusion, Emphysema, Enlarged Cardiomediastinum, Fibrosis, Fracture, Hernia, Infiltration, Lung Lesion, Lung Opacity, Mass, Nodule, Pleural Thickening, Pneumonia, Pneumothorax}.

\begin{figure}[!t]
\centering
\includegraphics[width=\columnwidth]{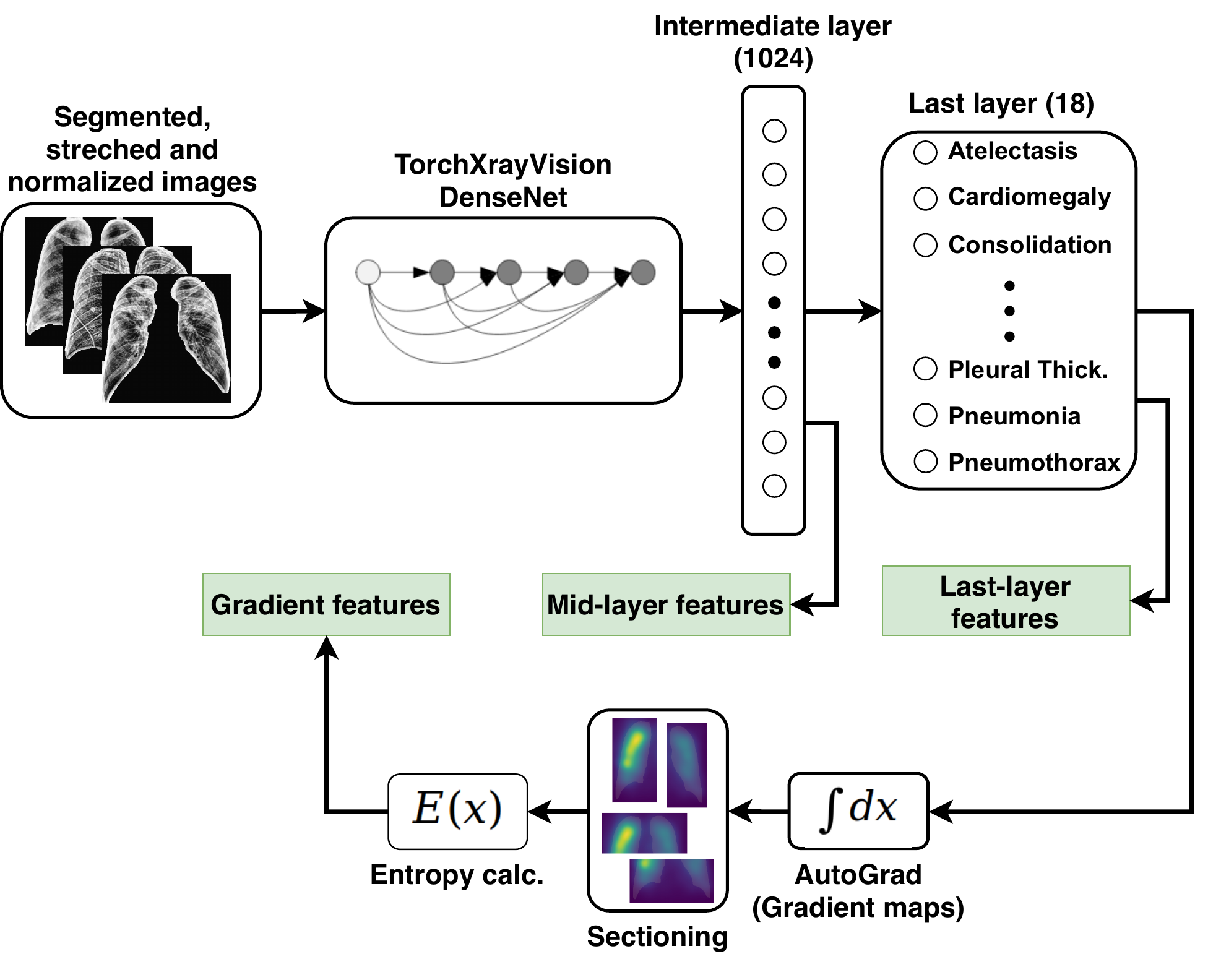}
\caption{Feature extraction pipeline. The segmented and normalized images are processed by a specialized pre-trained network. Features are extracted from parts of its architecture and hand-engineered from gradient maps.}
\label{feature_pipeline}
\end{figure}

Feature extraction from such a model is feasible from many parts of its architecture. In a way, any activation value from the forward pass through the network can be used as features. In this work, specifically, three modes were useful to attest separability between classes, assess the semantic meaning, and aid the pathology location. These methods were labelled ‘mid-layer’, ‘last-layer’, and ‘class-activation map’ features. The first method, extracting features from an intermediate layer, extracts the activation values from a layer previous to last, which contains 1024 nodes. These features were helpful to attest the separability between classes but, as they are not set to translate any semantic meaning, they were not used to infer any correlation with particular pathologies. The last-layer features were extracted from the activations of the 18-node layer, pre-sigmoid, and were useful at correlating some of the mentioned pathologies to the defined classes (ICU admission or not). The class-activation map features, or gradient features for short, were hand-engineered features soon to be described, set with the intention of aid the localization of the pathologies in the lung. Fig. \ref{feature_pipeline} is set to illustrate potential feature-extraction branches in the network architecture. It is worth mentioning that none of these features was combined or even used in their totality; they are presented as different classification scenarios where just a few are used, complying with the discussed initial intention to limit potential overfitting.   

\subsubsection{Gradient features}
\begin{figure}[t]
  \centering
  \begin{subfigure}[t]{0.8\columnwidth}
    \includegraphics[width=\textwidth]{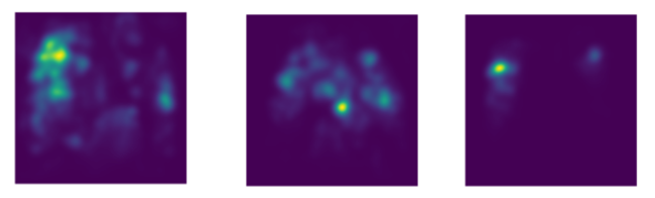}
    \caption{}
  \end{subfigure}             
  \begin{subfigure}[t]{0.8\columnwidth}
    \includegraphics[width=\textwidth]{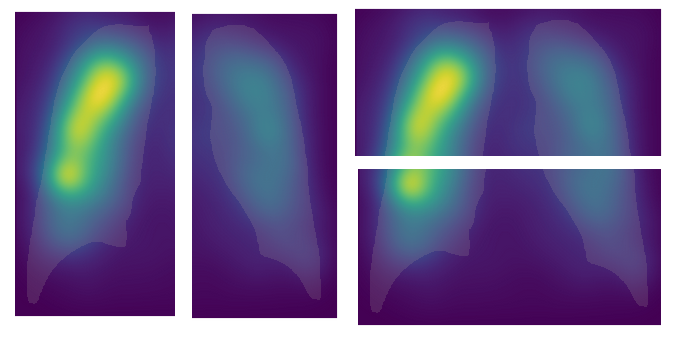}
    \caption{}
  \end{subfigure}
  \caption{Gradient features illustrations. a) Example of how the entropy metric changes; from left to right, entropy values: 2.92e-5, 6.77e-5, 14.19e-5 . b) Sectioning of gradient maps where the features are calculated: longitudinal (left) and transversal (right) cuts.}
  \label{entropy_feature}
\end{figure}

The mid-layer and last-layer features, despite sufficient when attesting separability, do not translate any information regarding the location of the pathologies in the lung. Therefore, gradient features were developed as a way to aid the correlation of the location of the lung pathologies with the patients with higher chances of developing severe symptoms. The calculation of the gradients in the network graph leaves in respect to an input image was performed by the autograd torch class, which is often used to create saliency maps and associate explainability to models \cite{gozes2020rapid,hu2020weakly,cohen2020predicting}. In this study, however, a procedure to create features from the gradient maps is proposed. First, the maps are construed with the energy of the accumulated gradients and then sectioned in two cuts: longitudinal and transversal. For each of these cuts, an entropy measure is calculated. The resulting feature is proposed here as a measure of the spread of the activations in their respective cuts. Such a single value measure is inspired by the Shannon entropy \cite{shannon2001mathematical} can be trivially calculated with Eq. \ref{eqn1}. Examples of such features and how the images were sectioned for spatial correlation are illustrated in Fig. \ref{entropy_feature}. 

\begin{equation} 
\label{eqn1} 
E = -\sum\limits_{j} \sum\limits_{i} p  \log p 
\end{equation}

\subsection{Classification and validation}
Given the advantages of interpretability and the commitment to limit the model complexity as primary factors in model selection, it is hard to argue for the application of another method than decision trees \cite{hastie2009elements}. Besides being human interpretable, one can easily limit the effects of overfitting by mechanisms of pruning, such as setting the minimum amount of samples in the tree leaves. For the experiments performed here, for example, trees were pruned to contain at least 20 samples in each leaf, so nodes are not branched in just a few samples. This constraint resulted in very shallow trees that only relied on a few features to make the classification (usually three to five). Such a pruning mechanism was the only hyper-parameter set when fitting the trees; the rest were all default values from the scikit learn package \cite{scikit}. It is worth noting here that the goal, despite trying to be as accurate as possible, is not claiming the ability to precisely predict the patients that will be admitted to ICU but discover potential semantic predictors that are correlated with such patients. 

The results are presented in three distinct scenarios with respective feature sets that the tree-fitting algorithms could greedily choose from mid-layer, last-layer, and gradient features. In each of these scenarios, the results are presented by fitting the tree in the whole data set and by performing cross-validation. To the latter, given the size of the data set, a simple 5-fold would probably not be representative of the predictive potential (especially when limited to leaf size of 10 samples). The number of folds was given by a leave-two-out approach where all the data set is used for training, except for 2 samples (one of each class), which are then used for testing. This procedure is repeated until all the samples in the dataset are used for testing. The shallow trees resulting from fitting the whole data set are also illustrated to depict their simple structure and interpretability. 

\subsection{Correlating results with an external data set}
In an attempt to add to the validation and address potential generalization concerns, a larger, external data set was also used in testing the fitted decision trees. 
Such a data set could only be used to attest correlations since its metadata does not contain labels regarding patient outcome (went to ICU or not).
The absence of these labels means that it could not be used for validation (in the strict sense), but the information it contains could be correlated to results from the learned semantic features.
This data set of X-ray images from COVID-19 patients, named BIMCV \cite{de2020bimcv} and publicly available, contains doctor’s annotations for each subject that were used here as discrete labels and descriptive terms.
Besides numerous, such labels refer not only to the pathologies observed by the doctors in the patient’s lungs but also points to their location.
Therefore, because they have a semantic meaning, they can then be correlated to the features from the last layer, which are also descriptive.

The method in which the features are correlated to the external data set is also presented as a contribution and, to the best of the authors' knowledge, not previously presented. As the quality of images in the data set varies a lot, the procedure starts by running the images through the same pre-processing pipeline described before with histogram equalization, lung segmentation and cropping, and feature extraction. Some images that did not have the lungs successfully segmented or were very low-quality are filtered out, but a total of 1312 images were labelled as valid. Subsequent to the processing pipeline, the decision tree performs the classification of features in the binary classes 0 and 1, and the samples pertaining to each class are assigned to two different sets. At this point, the correlation method enters the picture by creating a dictionary with the pathologies and locations labels and their respective frequency for each group. Such a frequency dictionary is then normalized by the set size and compared between classes. The correlation is a measure of how the features relevant to the classification in the first data set are represented (frequency) in the different classes assigned in the second data set. The null hypotheses, in this case, is that if the images were sampled at random, the frequency of labels would be equal in both sets. Such a procedure will become clearer in the presentation of the results in the following section.  

\section{Results}

Prior to discussing the classification metrics and feature work, it is worth noting the outcomes of training the lung segmentation network, also part of the contribution herein.
Trained on large chest X-rays image data sets, the designed architecture achieved a maximum validation dice similarity coefficient of 0.988 at epoch 93, which is comparable to the best result encountered thus far. 
The best performance observed in the literature was achieved using a complex CNN-based Deep Learning system, achieving a dice similarity coefficient of 0.980 \cite{hwang2017accurate}.
A much simpler approach leveraging a U-Net architecture \cite{ronneberger2015u}, similar to the the one here,  was trained on the JSRT dataset \cite{shiraishi2000development} consisting of 385 CXR images with gold-standard masks to achieve a dice similarity co-efficient of 0.974 \cite{novikov2018fully}.
Table \ref{lungseg} is set to contrast the performance of previously methods and the one proposed here.

\begin{table}[!t]
\renewcommand{\arraystretch}{1.5}
\caption{Previously proposed lung segmentation methods and respective dice similarity coefficient}
\label{lungseg}
\centering
\resizebox{.75\columnwidth}{!} {%
\begin{tabular}{cc}
\hline
Authors, citation & Dice similarity coef.\\
\hline
Li et al., \cite{li2016automatic} & 0.964 \\
Candemir et al., \cite{candemir2013lung} & 0.967 \\
Shao et al., \cite{shao2014hierarchical} & 0.972 \\
Novikov et al., \cite{novikov2018fully} & 0.974  \\
Yang et al., \cite{yang2017lung} & 0.975 \\
Hwang et al., \cite{hwang2017accurate}  & 0.980 \\
Ours & \bfseries{0.988} \\

\hline
\end{tabular}
}
\end{table}

\subsection{Separability}
The separability metrics were calculated for the final semantic features and other two different scenarios. The first refers to the non-task-specific features in the intermediate layer, which although contemplating 1024 nodes, had only 4 features used to attest invariance between classes. The second scenario regards the separability of the actual semantic (pathology labels) features output by the 18-node last layer of the network. The third, gradient features, will have its discussion delayed to a further sub-section regarding the pathology localizations, which is where its value really lies. The metrics presented as ‘whole set’ refers to using all samples to fit the tree while ‘cross-val’ indicates the ones resulting from the leave-two-out cross-validation approach. Such an approach means fitting the same number of trees as in the number of samples in the smallest class. For example, if one class has n samples and another has $m$, $n$ decision trees will be fitted if $n < m$.  

\begin{figure}[t]
  \centering
  \begin{subfigure}[t]{\columnwidth}
    \includegraphics[width=\textwidth]{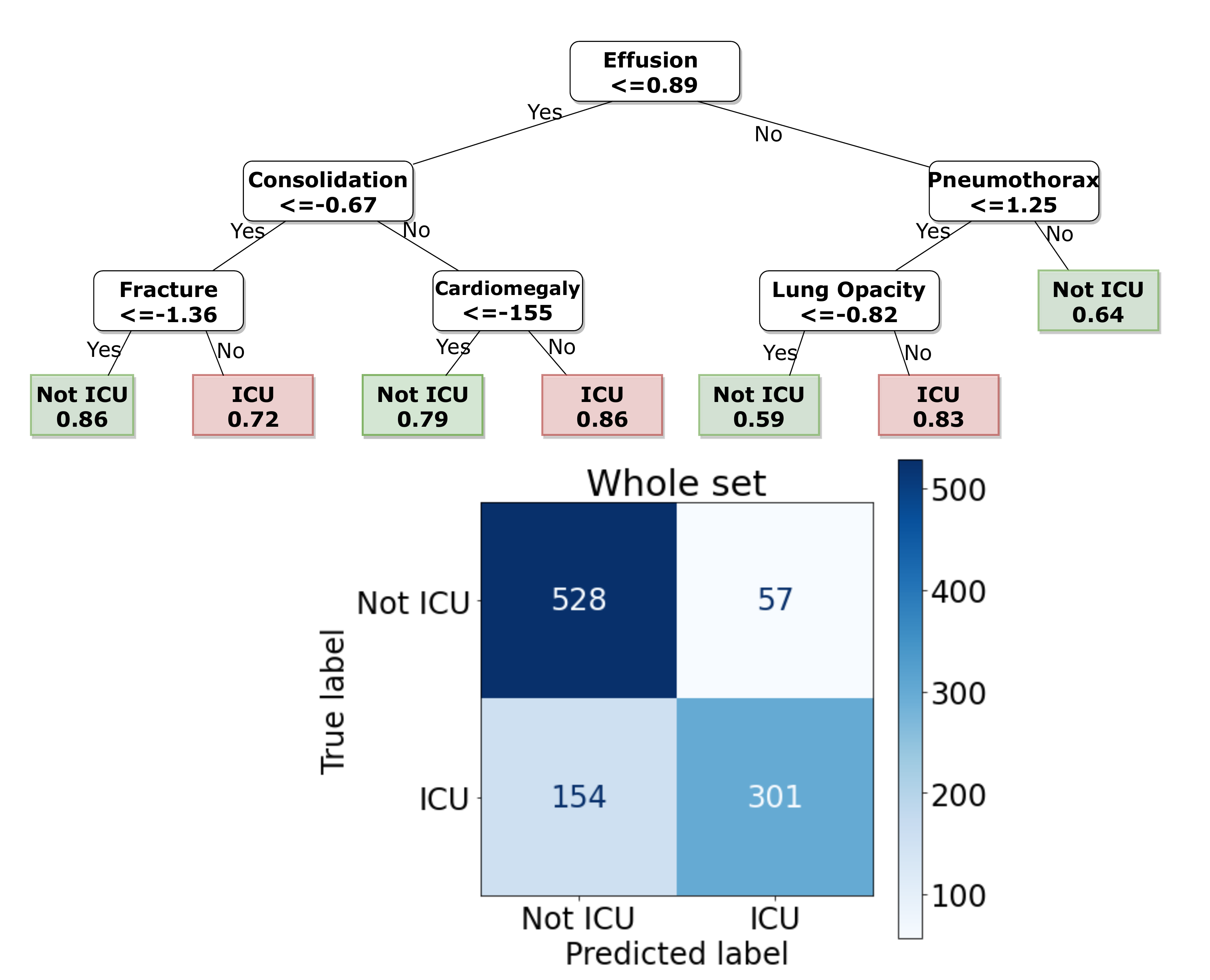}
    \caption{}
  \end{subfigure}             
  \begin{subfigure}[t]{\columnwidth}
    \includegraphics[width=\textwidth]{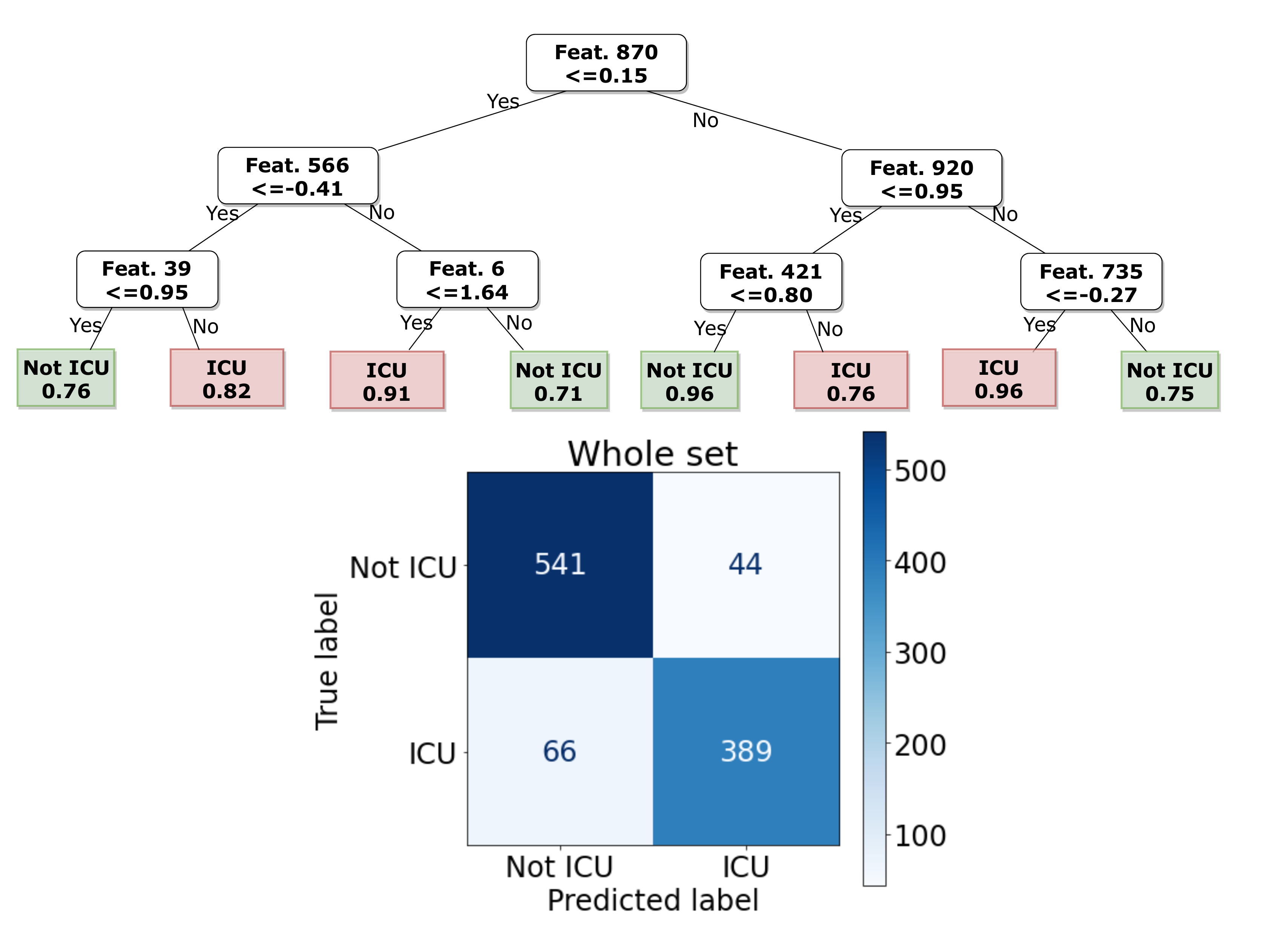}
    \caption{}
  \end{subfigure}
  \caption{Decision trees and confusion matrices from the three scenarios presented: a) last-layer and b) mid-layer features.}
  \label{decision_trees1}
\end{figure}

The resulting metrics showed that the mid-layer features are better at separating the classes but not for a large margin when compared to the semantic features from the last layer. The mid-layer whole set features resulted in 0.89 accuracy and 0.88 F1-score, while the last-layer features' accuracy was 0.80 with a 0.74 F1-score. Such metrics were obtained when limiting the minimum leaf size to 20 samples with a maximum tree depth of 4; both values were arbitrarily set by trial and error attempts guided by cross-validation. 
The whole-set tree is the result to be highlighted here as it has allegedly semantic features relating to the lung pathologies. 
If the features are ranked by samples under their respective node, it can result in interesting insights since they are trained to have semantic meaning. 
The three features with most samples, in this case, were `Effusion', `Consolidation',  and `Cardiomegaly'.
Fig. \ref{decision_trees1} depicts a 3-level version of the tree for better illustration purposes, which has performance closely equivalent to the 4-level version.
Such labels will be important in the following subsection when correlating with the external dataset and in the discussion section where they will be qualitative compared to the pathologies referred to in the literature regarding patients that develop severe symptoms. As for the cross-val scenario, mid-layers and last-layer features resulted in 0.86 and 0.76 accuracies, respectively. The distribution of predicted versus real labels can be inspected by the decision matrices presented in Fig. \ref{decision_trees1}. As an additional worth-mentioning note from preliminary experiments, it was observed that such accuracies would drop to values close to 0.5 if the normalization techniques were not applied, which was significant evidence of their importance. 

\subsection{Correlating with an external data set}
The many labels contained in the BIMCV data set allowed for an interesting comparison between the classes after classification was applied to the images. From the 1312 images selected after filtering, 73 did not have text labels assigned, but most had at least a few pathology and location labels. After the images went through the processing and feature extraction pipeline, the fitted decision tree (whole set, last layer) classified the features and their respective text labels were assigned to sets given by their predicted class. In this manner, two dictionaries could be construed with the keys being the unique words in the set, and the values representing the frequency of such labels. 
The five most frequent pathological labels in the data set can be listed as examples of the types of terms describes:  ‘pattern’ (467), ‘COVID’ (443), ‘pneumonia’ (301), ‘increased density’ (368) and ‘unchanged’ (272). 

The comparison presented here was set to be the normalized ratio between the frequency of words in the potentially severe class (1) and not (0). By ‘normalized’ ratio, it is meant that the frequency of words was multiplied by the inverse of the number of samples in the respective set before the ratio calculation. The number of samples in each set after classification was 806 images in class 0 and 506 in class 1. The ratio was calculated only for words that had the minimal arbitrary number of appearances of 20 in each class. It may be worth noting that, if the samples were picked at random, despite the selection of images before classification or the number of images in a set, such a ratio would be equal to 1, given the normalization. 
Table \ref{table1} presents the pathological and localization labels with normalized ratios higher than 1.2 and lower than 0.8, in descending order.
The most interesting aspect from such comparison is certainly the fact that two of the most over-represented features in the images of the external data set were also the ones chosen as most important when fitting the decision tree in the data set used for training. 
This shows that the features used to separate the data set used for training are also consistently detected on the external data set.
Moreover, two over-represented features in the class of higher chances to go severe — ‘Consolidation’ and ‘Bilateral’ — are often cited as pathologies relating to the severity in the literature \cite{ bernheim2020chest, zhao2020relation}. 
It is also pointed out that a systematic review on the subject showed that the prevalent locations were bilateral and peripheral (another over-represented term from results presented here).

\begin{table*}[!t]
\renewcommand{\arraystretch}{1.2}
\caption{Over- and under-represented pathology and localization labels and their frequency ratio }
\label{table1}
\centering
\resizebox{1.3\columnwidth}{!} {%
\begin{tabular}{cccc}
\multicolumn{2}{c}{Pathology} & \multicolumn{2}{c}{Localization} \\
\hline
Feature & Ratio (C1/C0) & Feature & Ratio (C1/C0)\\
\hline
\bfseries{‘Consolidation’} (127)  & 1.67 & \bfseries{‘Bilateral’} (449)  & 1.58\\
‘Alveolar’ (151)  & 1.34 & ‘Middle’ (326)  & 1.39\\
\bfseries{‘Effusion’} (60) & 1.30 & ‘Lower’ (452)   & 1.31\\
‘COVID’ (443)  & 1.24 & \bfseries{‘Peripheral’} (489) & 1.28\\
‘Pneumonia’ (301)  & 1.24 & ‘Upper’ (305) & 1.25\\
‘Pleural’ (67) & 1.22 & ‘Left’ (549) & 1.23\\
‘Infiltrates’ (184)  & 1.23 & ... & {} \\
‘Interstitial’ (223)  & 1.2 & ‘Hilar’ (99)   & 0.66\\
... & {} & ‘Mediastinum’ (84)  & 0.56\\
\bfseries{‘Normal’} (168)  & 0.48 & {} & {} \\

\hline
\end{tabular}
}
\end{table*}

\subsection{Spatial analysis and visualization}
While the pathological features presented in the previous subsections are expressive and correlate with the frequency of labels in the external data set, they cannot be correlated to any information regarding the locations of the pathologies. The features are scalars given by their respective layer’s activation function, which does not translate information about their spatial distribution. The simple method proposed here uses hand-engineered features on gradient maps to check for areas where the predictive features are prevalent. The decision trees were trained in the same manner as other features but, in this case, only a two-level tree was used to aggressively limit overfitting. Every feature relates to an entropy metric in one of the segments extracted from the gradient maps: left and right longitudinal, and superior and inferior transversal cuts (see Fig. \ref{entropy_feature}). 

\begin{figure}[!t]
\centering
\includegraphics[width=\columnwidth]{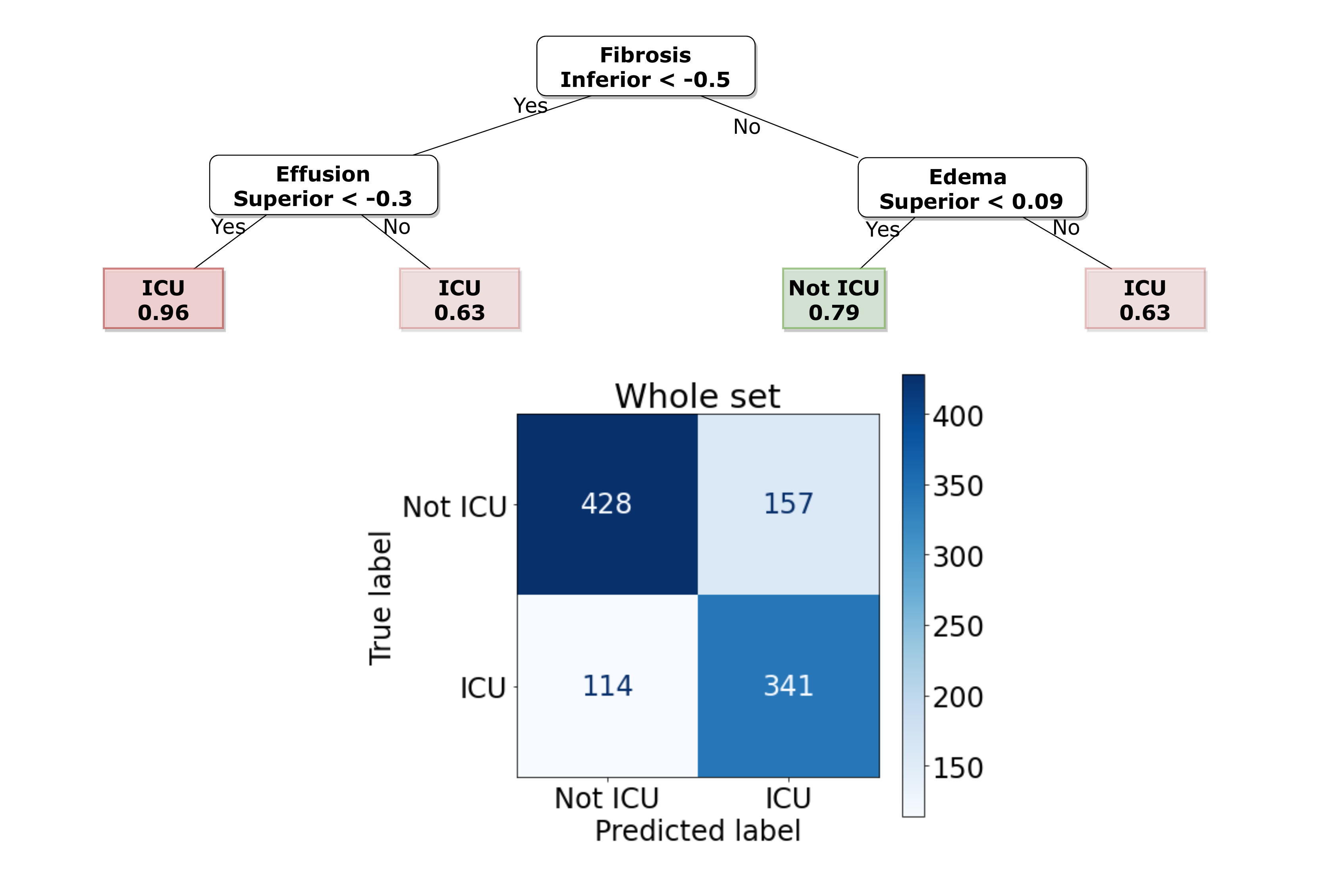}
\caption{Decision tree and confusion matrix from the entropy gradient features.}
\label{decision_trees3}
\end{figure}

The entropy features extracted from the gradient maps segments and selected as splits in the decision tree presented some correlation with the external data set and other reported findings. Regarding the separability metrics, fitting a shallow decision tree with only three features in the whole data set resulted in 74\% accuracy (separability), while with 72\% accuracy score in cross-validation. Shown in Fig. \ref{decision_trees3}, the three features were 
‘Fibrosis’ on the Inferior, and ‘Effusion’ and ‘Edema’ on the Superior cut.  
Although these features don't translate any definitive information, such an interpretable method shows to be a promising way to analyse the spatial relationships between features and potential severity. 

\begin{figure}[t]
  \centering
  \begin{subfigure}[t]{0.9\columnwidth}
    \includegraphics[width=\textwidth]{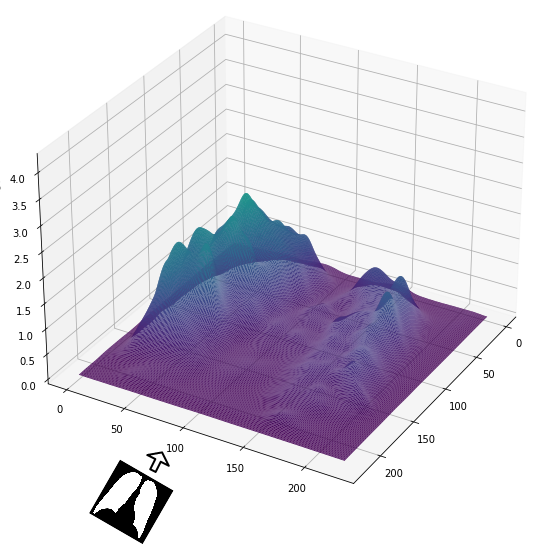}
    \caption{}
  \end{subfigure}             
  \begin{subfigure}[t]{0.9\columnwidth}
    \includegraphics[width=\textwidth]{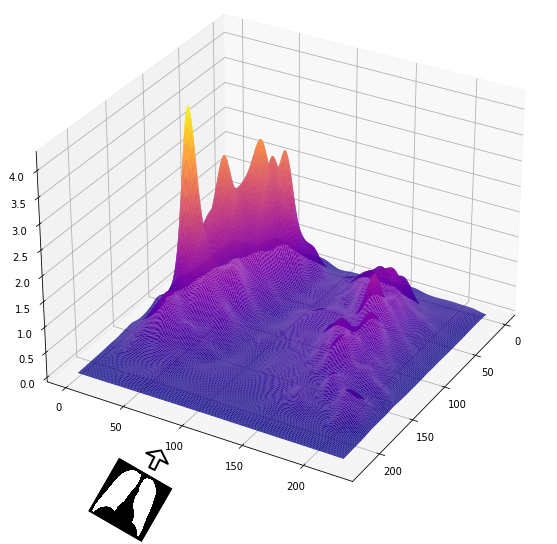}
    \caption{}
  \end{subfigure}
  \caption{Surface plots of the average of gradient maps in each class: a) Non-ICU and b) ICU.}
  \label{heatmaps}
\end{figure}

Due to the fact that the methodology includes the segmentation and cropping of the lung area as an important normalization measure, an interesting way to visualize the spatial distribution of some features becomes possible. Such a way proposed here differs from the usual by not just plotting the energy of the gradient maps of a particular feature but averaging all the maps of a certain class and then platting that as a surface. An example is presented in Fig. \ref{heatmaps} where the chosen feature is the one in the top of the tree of semantic features: ‘Effusion’. The two images are the average of activations of the two classes, normalized by class size. The plotted surfaces show that such a feature not only has gradients with higher energy in the images but also are more spatially distributed throughout the lung (smaller entropy). 

\section{Discussions}
It is worth iterating that the contribution here is not the development of a deployment-ready classifier but the presentation of pathological features with similar characteristics as ones observed in patients with higher chances of developing severe symptoms. 
As works in the literature have been discussing \cite{cohen2020covidProspective, rubin2020role, wynants2020prediction}, using Deep Learning methods in clinical settings is highly inadvisable, even when interpretable. 
Practical prediction claims would already be restrained by the classification metrics presented here, but they still can’t take away the fact that some features showed interesting correlations and significance to patient severity.
Moreover, the contribution here is not constrained to the predictive potential of features; it also includes the methodology developed to evaluate and select them.
More than informing practitioners of evidence for setting heavier weights to the importance of some pathologies in disease development, one can hope that the method presented could be useful in for problems in similar domains. 
Given that computer vision-based COVID-19 diagnostic tools are not recommended for practical use due to their potential bias and risk \cite{cohen2020covidProspective, rubin2020role, wynants2020prediction}, developing systems that can support human decisions in critical resource management becomes one of the most promising uses of Machine Learning \cite{cohen2020covidProspective}. 

Regarding significance, it is arguable that the most significant aspect of the results is how the decision tree, fitted by a limited dataset, selected images on an external data set that had an over-representation of labels relating to disease severity. 
The two most over-represented features in the class of higher chances to go severe — ‘Consolidation’ and ‘Bilateral’ — are often cited as pathologies relating to severity in the literature \cite{yoon2020chest, bernheim2020chest, zhao2020relation}. 
For example, the authors in \cite{ng2020imaging} reported that the evolution from ground-glass opacities to consolidation was present in some severe patients. 
They also pointed out that a systematic review on the subject showed that the prevalent locations were bilateral and peripheral (another over-represented term from results presented here).
Another highly cited work \cite{huang2020clinical} reported that most of the patients had bilateral involvements and that ICU-admitted patients showed bilateral multiple lobular and subsegmental areas of consolidation.

\begin{figure}[!t]
\centering
\includegraphics[width=\columnwidth]{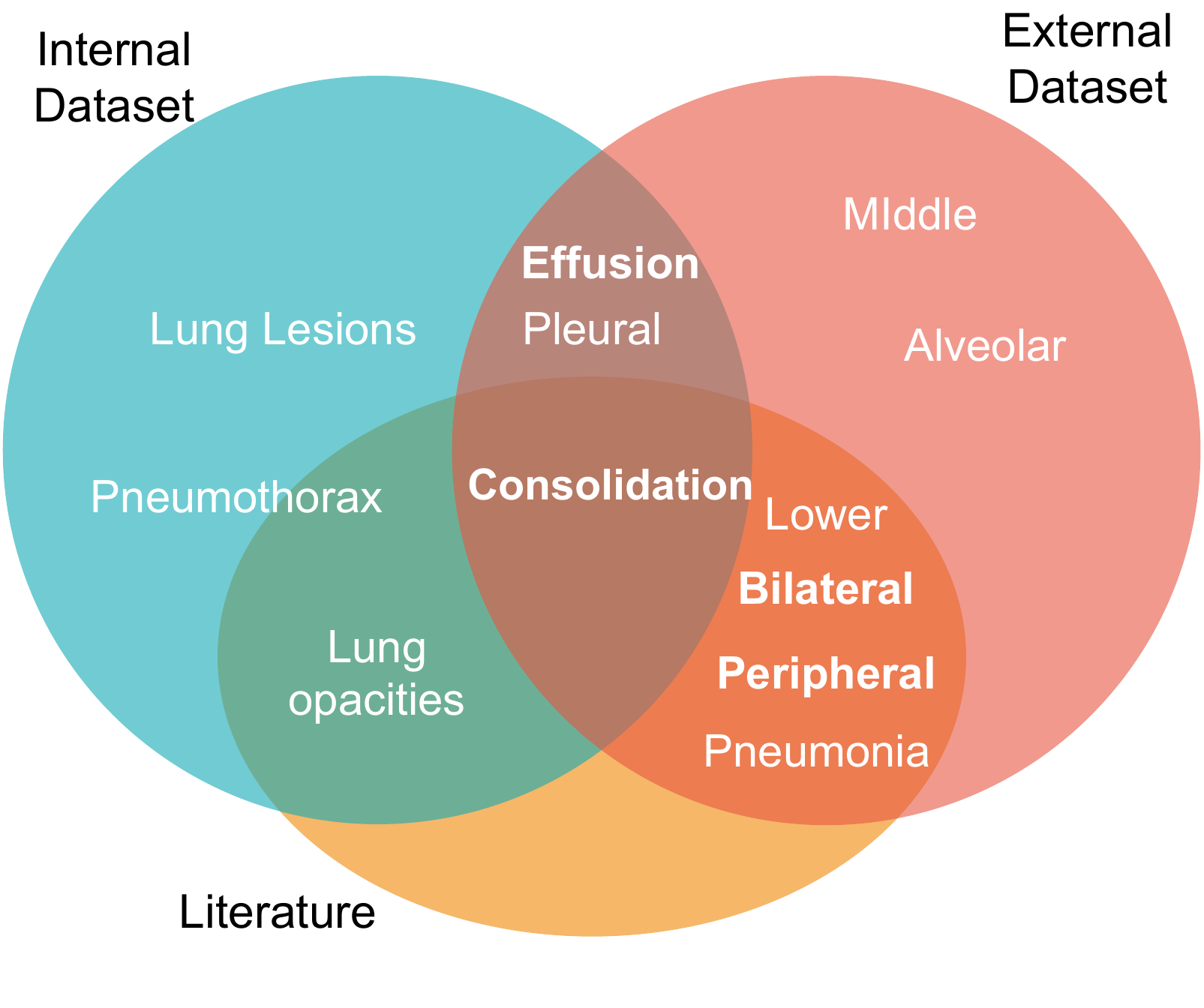}
\caption{Venn diagram illustrating some of the correlated findings from the internal, external, and literature reported pathologies and locations.}
\label{venn_diag}
\end{figure}

Nevertheless, one should also note that these pathologies were not present in all severe patients, neither in the analysis presented here and in the literature.  
This observation should be evident since lung pathologies are not the only factor in determining severity of COVID-19 and ICU admission.
Co-morbidities and other systemic involvement in COVID-19 also play an important role in this disease, thus representing the primary limitation of the present work. 
Fig. \ref{venn_diag} summarizes and illustrates some of the worth-noting correlations found here are for improved clarity and comprehension.

\section{Conclusions}
The results presented in this paper point to the fact that some semantic features (presented here as pathology scores) have reasonable predicting potential to patients that eventually develop severe symptoms and are admitted to ICU. Extracted from a limited data set of chest x-ray images, the features only presented any predictive potential when enough care was taken to prevent overfitting. Such measures included training a state-of-the-art lung segmentation network, using a pre-trained, specialized network to extract the features, image normalization techniques, and strictly selecting an interpretable classifier of low complexity. Results from running the fitted classifier on an external data set showed that it indeed selects for the semantic features chosen when learning. These features not only appeared more frequently in the set of patients classified as with higher chances of going severe but a frequency analysis on all labels showed that descriptors often used to describe COVID-19 pathologies were over-represented in such patients. 
This observation was also attested in localizations labels, which showed over-representation of important terms like ‘bilateral’, ‘peripheral’, and ‘middle’. 
Such a methodology for creating correlations between interpretable features and patient outcome may aid human decision-making in resource management and be a template method for similar problems from other domains. 
More results, analyses, and source code of experiments can be found in the in work repository \footnote{\url{https://github.com/dougpsg/covid_mavidh_icufeatures_scoring}}.



%





\ifCLASSOPTIONcaptionsoff
  \newpage
\fi



\bibliographystyle{IEEEtran}
\bibliography{references.bib}
%








\end{document}